\documentclass[twocolumn,showpacs,preprintnumbers,amsmath,amssymb]{revtex4-1}

\usepackage[dvipdfmx]{graphicx}
\usepackage{dcolumn}
\usepackage{bm}

\begin{document}

\preprint{APS/123-QED}

\title{Pudding-mold type band in a potential thermoelectric 
material CuAlO$_2$ : comparison with Na$_x$CoO$_2$}

\author{Kouta Mori$^{1,2}$}
\author{Hirofumi Sakakibara$^1$}
\author{Hidetomo Usui$^{1,2}$}
\author{Kazuhiko Kuroki$^{2,3}$}

\affiliation{$\rm ^1$Department of Engineering Science, The University of Electro-Communications, Chofu, Tokyo 182-8585, Japan}
\affiliation{$^2$ JST, ALCA, Gobancho, Chiyoda, Tokyo 102-0076, Japan}
\affiliation{$^3$ Department of Physics, Osaka University, 
1-1 Machikaneyama, Toyonaka, Osaka 560-0043, Japan}

\date{\today}

\begin{abstract}
A potential thermoelectric material CuAlO$_2$ is theoretically studied. 
We first construct a model Hamiltonian of 
CuAlO$_2$ based on the first principles band calculation, and 
calculate the Seebeck coefficient. Then, we compare the model with that 
of a well-known thermoelectric material Na$_x$CoO$_2$, 
and discuss the similarities and the differences.
It is found that the two materials are similar 
from an electronic structure viewpoint in that 
they have a peculiar pudding-mold type band shape, which 
is advantageous as thermoelectric materials.
There are however some differences,  
and we analyze the origin of the difference from a microscopic 
viewpoint. 
The band shape of CuAlO$_2$ is found to be 
even more ideal than that of Na$_x$CoO$_2$, and  
we predict that once a significant amount of holes is doped in 
CuAlO$_2$, thermoelectric properties (especially the power factor) 
even better than that of Na$_x$CoO$_2$ can be expected.
\end{abstract}

\maketitle

\section{INTRODUCTION}
Search for good thermoelectric materials serves as an intriguing challenge 
both from the viewpoint of fundamental physics as well as device 
applications\cite{Mahan}.
Thermoelectric materials are often found in semiconductors,  but the 
discovery of a large Seebeck coefficient found in Na$_x$CoO$_2$\cite{Terasaki} 
which exhibits a metallic nature of the conductivity opened
up a new avenue for the search of thermoelectric materials.
Coexistence of the large Seebeck coefficient $S$ and the low resistivity $\rho$
can give rise to a large power $P=S^2/\rho$, which is important from a 
application point of view.
Soon after the discovery, the origin of the large thermopower in 
Na$_x$CoO$_2$ was studied theoretically, where the possible importance of the 
orbital degeneracy 
\cite{Koshibae,KoshibaePRL} 
or the narrow band width\cite{Singh} has been pointed out.
Later on, one of the present authors along with Arita 
proposed that the peculiar band shape 
in which the top is flat but 
bends sharply into a dispersive portion 
plays an important role in the 
coexistence of the large Seebeck coefficient and the 
low resistivity in Na$_x$CoO$_2$\cite{Kuroki}. 
This band has been referred to as the ``pudding-mold'' type.
Recently, various materials with good thermoelectric 
properties, such as CuRhO$_2$\cite{Kuriyama,Shibasaki2,Usui}, 
Li$_2$RhO$_4$\cite{Okamoto,Arita}, 
and FeAs$_2$\cite{PSun,UsuiFeAs} have been 
shown to possess this type of band shape. The effect of the 
interplay between electron correlation and 
this kind of band shape  has also been studied
\cite{Arita,Wissgott,Wissgott2,UsuiNaxCoO2}.

Along this line of study, here we focus on CuAlO$_2$.
In fact, several previous studies have pointed out a strong potential 
of this material as a good thermoelectric material\cite{Yoshida,Poopanya}, 
although sufficient 
carrier doping to reduce the resistivity has not been successfully accomplished 
so far. In ref.\cite{Yoshida,Nakanishi}, band structure calculation 
has been performed, which shows a flat portion at the top of the band, 
which is reminiscent of the band shape of Na$_x$CoO$_2$.
In the present paper, we first construct a model Hamiltonian of 
CuAlO$_2$ based on the first principles band calculation, and 
calculate the Seebeck coefficient. Then, we compare the model with that 
of Na$_x$CoO$_2$, and discuss the similarities and the differences.
We also analyze the origin of the difference of the band structure 
between the two materials from a microscopic point of view. 
The pudding-mold type band shape of CuAlO$_2$ is found to be 
even more ideal than that of Na$_x$CoO$_2$, and  
we predict that once a significant amount of holes is doped in 
CuAlO$_2$, thermoelectric properties (especially the power factor) 
even better than that of Na$_x$CoO$_2$ can be expected.
This is due to the combination of an extremely ideal band shape 
(enhances the Seebeck coefficient) and a relatively 
large band width (can give large conductivity when holes are doped).

\begin{figure}[htbp]
\includegraphics[width=6cm]{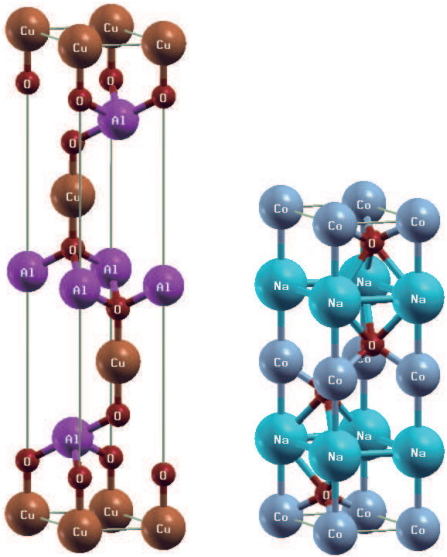}
\caption{The lattice structure of CuAlO$_2$(left) and Na$_x$CoO$_2$(right)}
\label{fig1}
\end{figure}

\section{Construction of the  single orbital model}
\begin{figure}[htbp]
\includegraphics[width=8cm]{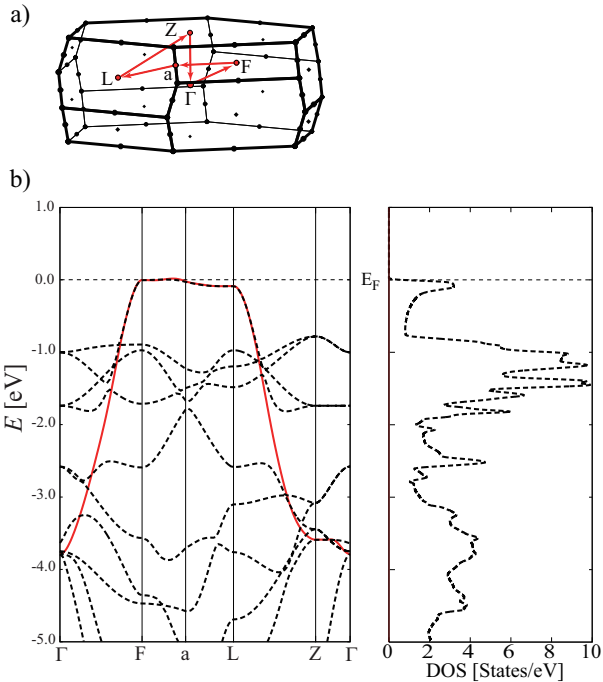}
\caption{(a)The Brillouin zone (b)(left) The first principles band calculation 
result of CuAlO$_2$(dashed lines) and the single orbital model (solid line) 
(right) the density of states.}
\label{fig2}
\end{figure}

We perform first principles band calculation of CuAlO$_2$ 
using the Wien2k package\cite{wien2k}. We adopt the lattice structure 
parameters given in ref.\cite{Nakanishi}.
Here we take $RK_{\rm max}=7$, 512 $k$-points, and 
adopt the PBE-GGA exchange correlation functional\cite{PBE}.
The calculation result is shown as dashed lines in Fig.\ref{fig2}, which  
is essentially the same as those obtained in previous 
studies\cite{Yoshida,Nakanishi}. Namely, there is a flat portion around the 
Brillouin zone edge, which bends sharply into a dispersive portion 
as the $\Gamma$ point is approached. This is nothing but the pudding-mold 
type band introduced in ref.\cite{Kuroki}.

Since the flat portion of the band is well isolated from other portions, 
we can extract this portion and construct a single orbital model, where 
we exploit the maximally localized Wannier orbitals\cite{wannier,w2w}. 
Here, the extracted orbital 
is obtained by projecting onto the $3d_{3z^2-r^2}$ orbital, namely, the 
obtained Wannier 
orbital has strong Cu $3d_{3z^2-r^2}$ character, while the contributions 
from other orbitals due to the hybridization are 
implicitly taken into account in this single Wannier orbital.
The result is shown in Fig.\ref{fig2} as a solid 
line superposed to the first principles result.

\section{Seebeck coefficient and Power factor}

\subsection{The Boltzmann's equation approach}
In this section we calculate the Seebeck coefficient adopting the 
single orbital model and using the Boltzmann's equation approach, 
In this approach, 
tensors of the Seebeck coefficient $\bm{S}$ and the 
conductivity  $\bm{\sigma}$ are given as, 
\begin{equation}
\bm{S}=\frac{1}{eT}\bm{K}_1 \bm{K}_0^{-1} \label{Seebeck}
\end{equation}
\begin{equation}
\bm{\sigma}=e^2\bm{K}_0 \label{EC}
\end{equation}
where $e(<0)$ is the elementary charge, $T$ is the temperature、tensors 
$\bm{K}_1$,$\bm{K}_2$ are given as 
\begin{equation}
\bm{K}_n=\sum_{\bm{k}}\tau(\bm{k})\bm{v}(\bm{k})\bm{v}(\bm{k})\left[-\frac{\mathrm{d}f(\epsilon)}{\mathrm{d}\epsilon}(\bm{k})\right]\left(\epsilon(\bm{k})-\mu\right)^n \label{K0K1}.
\end{equation}
Here, $\epsilon(\bm{k})$ is the band dispersion,
$\bm{v}(\bm{k})=\frac{1}{\hbar}\nabla_{\bm{k}}\epsilon(\bm{k})$ is the 
group velocity, $\tau(\bm{k})$ is the quasiparticle lifetime, 
$f(\epsilon)$ is the Fermi distribution function, and $\mu$ is the 
Fermi level (chemical potential).
Here, due to the derivative of the 
Fermi distribution function $\mathrm{d}f(\epsilon)/\mathrm{d}\epsilon$, 
large contributions to $K_0$ and $K_1$ come from within $k_BT$ from 
the Fermi level $\mu$. 
In the present study, we approximate $\tau$ as a constant, so that 
it cancels out in the Seebeck coefficient. We simply write 
$\sigma_{xx}$ and $S_{xx}$ as $\sigma$ and $S$, respectively.
$\sigma$ and thus the power factor ${\sigma}S^2$ contains the 
constant $\tau$, whose absolute value is not determined.
Therefore, we only discuss the values of the power factor 
normalized by its maximum value as a function of hole doping rate.

\subsection{The pudding mold type band}
As discussed in ref.\cite{Kuroki}, pudding-mold type band is 
advantageous in obtaining large Seebeck coefficient despite 
low resistivity.
For a constant $\tau$, eq. (\ref{K0K1}) can roughly be 
approximated as 
\begin{equation}
\begin{cases}
K_0=\tau\sum_{k}\left( v_{\mathrm{above}}^2 + v_{\mathrm{below}}^2 \right)  \\
K_1=\tau\sum_{k}\left( v_{\mathrm{above}}^2 - v_{\mathrm{below}}^2 \right), 
\end{cases}\label{apK0K1}
\end{equation}
where $v_{\mathrm{above}}$ and $v_{\mathrm{below}}$ are group velocities above and 
below the Fermi level (representative values within $k_BT$ from the 
Fermi level). Since the Seebeck coefficient is proportional to 
${K_1}/{K_0}$, a larger difference between $v_{\mathrm{above}}$ and 
$v_{\mathrm{below}}$ gives larger $S$. Physically, this means that a 
large difference in the group velocities of holes and electrons results in a 
large Seebeck coefficient. When the Fermi level lies near the band edge, 
this difference 
(the $v_{\mathrm{below}}/v_{\mathrm{above}}$ ratio) 
can be large but the absolute values of the 
velocities are small, so that the conductivity becomes small.
On the other hand, in usual metallic systems, in which the 
Fermi level lies in the middle of the bands where 
$v_{\mathrm{above}}$ and $v_{\mathrm{below}}$ have similar values, the Seebeck 
coefficient tends to be small. This is the reason why a large power 
factor ${\sigma}S^2$ is usually difficult to obtain.
For the pudding-mold type band, however, 
the large density of states at the top of the band prevents the 
Fermi level from going down rapidly even when a large amount of 
carriers (holes in the present case) is doped, and when the 
Fermi level sits close to the bending point of the band, 
$K_1$ is large because of the small  $v_{\mathrm{above}}$ due to the 
flat portion and the large $v_{\mathrm{below}}$ due to the dispersive portion
of the band. In this manner, the coexistence of large Seebeck 
coefficient and small resistivity is realized for a wide range of hole 
doping ratio.

\subsection{Calculation Results}
\begin{figure}[htbp]
\includegraphics[width=8cm]{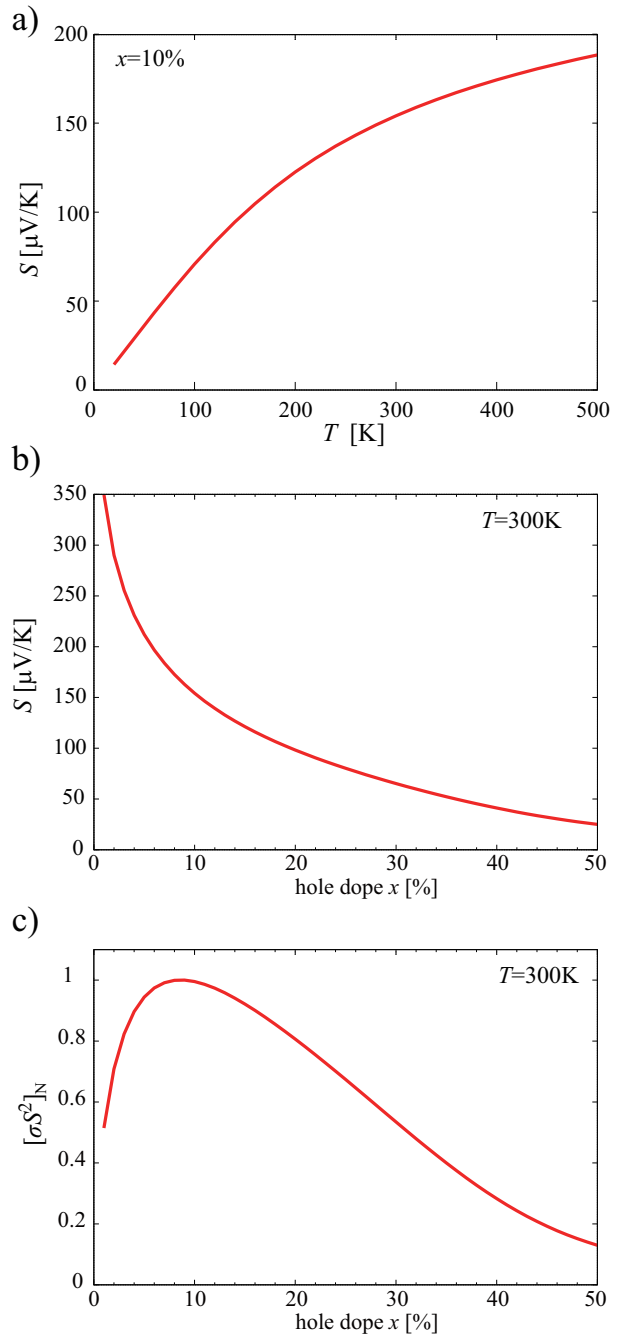}
\caption{The calculated thermoelectric properties of CuAlO$_2$. (a) The Seebeck coefficient against temperature for 10\% hole doping (b) The Seebeck coefficient against the hole doping concentration at $T=300$K. (c) The normalized power 
factor against the doping concentration at $T=300$K.}
\label{fig3}
\end{figure}

Fig.\ref{fig3} shows the Seebeck coefficient calculated for the 
single orbital model of CuAlO$_2$. Fig.\ref{fig3}(a) 
shows the temperature dependence for 10\%  hole doping, 
and Fig.\ref{fig3}(b)  is the doping dependence at $T=300$K.
These calculated values (e.g. $150\mu$V/K at $T=300$K for 10\% doping)
are similar to those values calculated for the single band model of 
Na$_x$CoO$_2$ in ref.\cite{Kuroki}. It is worth mentioning that 
in ref.\cite{Kuroki}, the band width was reduced to $\simeq$ 1eV (by hand)  
so as to fit the angle resolved photoemission data, while in the 
present case the  band width is about 4eV, i.e., four times larger.
A larger band width implies a larger group velocity in the 
dispersive portion of the band, which can give rise to better conductivity.
The Seebeck coefficient being nearly the same despite the much larger 
band width implies that the band {\it shape} of CuAlO$_2$ is even more 
ideal than that of Na$_x$CoO$_2$ from the viewpoint of obtaining large 
thermopower. Namely, the group velocity of the electrons 
is ideally close to zero (the band above the Fermi level is very flat), 
so that only the holes contribute to the thermopower.

The maximum power factor is reached around 10 percent hole doping, 
which is a rather large doping rate, and is in fact similar to the situation in 
Na$_x$CoO$_2$\cite{Kuroki}. This is another feature peculiar to the 
pudding-mold type band, where  a large amount of doping does not 
result in a rapid reduction of the Fermi level.

\section{Origin of the pudding-mold type band}
\subsection{The effect of the 2nd and the 3rd nearest neighbor hoppings}
\begin{figure}[htbp]
\includegraphics[width=8cm]{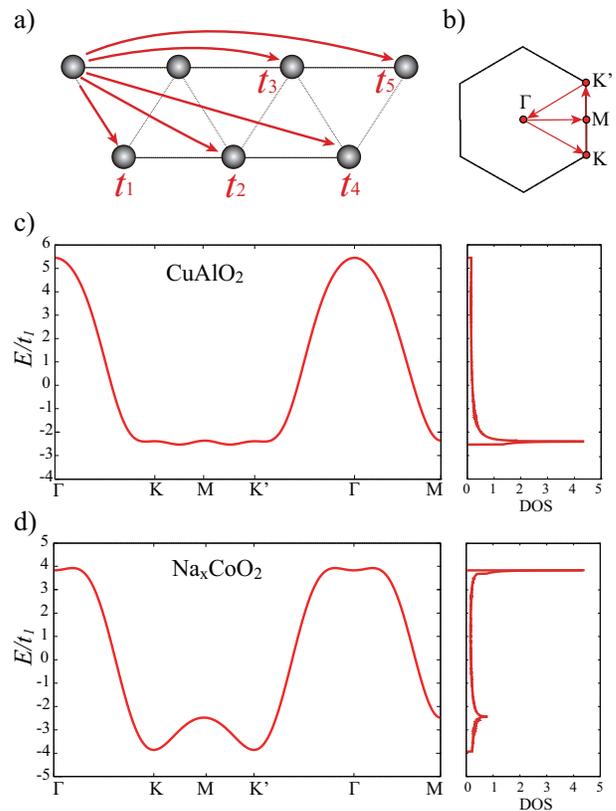}
\caption{(a)The definition of the hoppings on the 
triangular lattice (b) the two dimensional Brillouin zone, 
(c)(d) band dispersion 
normalized by $t_1$ and the density of states 
of the two dimensional 
models of CuAlO$_2$ and Na$_x$CoO$_2$ which considers the 
extracted hoppings $t_1\sim t_5$.}
\label{fig4}
\end{figure}

Both CuAlO$_2$ and Na$_x$CoO$_2$ exhibit pudding-mold type band.
Here we discuss its origin from the viewpoint of the hopping integrals 
on the triangular lattice.
Here we focus on the band structure within the planes,  
and neglect the hopping integral in the $z$ direction.
In Fig.\ref{fig4}, we show the band structure of the single band 
model of both materials 
in the Brillouin zone of the two dimensional triangular lattice,
in which the hoppings up to fifth nearest neighbors are extracted 
($t_1\sim t_5$, the original model contains very small hoppings between 
more distant sites). 
The band structure is normalized by the nearest neighbor 
hopping $t_1$. Note that the band structure of CuAlO$_2$ is turned 
up-side down compared to the original one  because $t_1<0$.
The hoppings normalized by $t_1$ are given in table \ref{table1}.
From this comparison, it can be seen that the flat portion of the 
band occurs around the Brillouin zone edge in CuAlO$_2$, while 
the flat portion of the band structure of Na$_x$CoO$_2$ is around the 
$\Gamma$ point. Note that the path along the Brillouin zone edge 
-K-M- in the 
two dimensional Brillouin zone corresponds to lines like -a-L(or F)- 
in the three dimensional Brillouin zone shown in Fig.\ref{fig2}(a).

\begin{figure}[htbp]
\includegraphics[width=8cm]{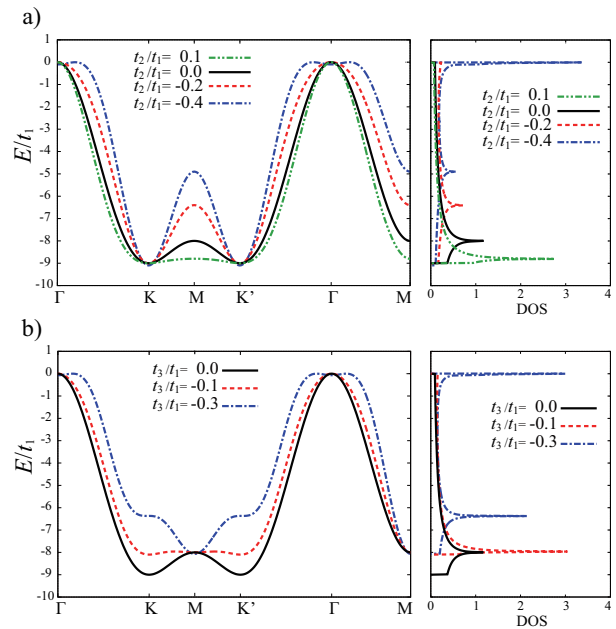}
\caption{Band dispersion of the single orbital model on 
the triangular lattice in which only $t_1$ and $t_2$(a) or $t_3$(b) are 
considered.}
\label{fig5}
\end{figure}

To see the origin of this difference between the two materials, 
we vary the hopping integrals $t_2$ and $t_3$ ``by hand''. 
As shown in Fig.\ref{fig5}, the second nearest neighbor 
hopping has a dramatic effect of making the band around 
the $\Gamma$ point flat while making it around K-M dispersive.
Consequently, the peak structure of the density of states moves 
largely from $E<0$ to $E>0$ as the absolute value of $t_2<0$ is increased.
The third nearest neighbor hopping $t_3<0$ on the 
other hand has the effect of  reversing the dispersion 
around K-M, so that the band around K-M once becomes 
nearly perfectly flat (around $t_3/t_1=-0.1$) as $|t_3|$ is increased.

Now, if we go back to the hopping integrals of the two materials
given in table\ref{table1}, 
there is a large difference, i.e.,  $|t_2/t_1|$ is almost 
negligible in CuAlO$_2$ compared to that in Na$_x$CoO$_2$.
This, along with $t_3/t_1$ being somewhat close to $-0.1$, makes the 
band around K-M fairly  flat in CuAlO$_2$.
Strictly speaking, $t_3/t_1\sim -0.07$ is 
still somewhat away from $t_3/t_1=-0.1$, where a nearly perfectly 
flat band appears. Actually, we find that there is also the effect of 
$t_5/t_1>0$, which makes the band around K-M flat, while 
enhancing the dispersion around the $\Gamma$ point.
In sharp contrast to CuAlO$_2$, the combined effect of $t_2$ and $t_3$ in 
Na$_x$CoO$_2$  makes the band around the $\Gamma$ point flat, 
while enhancing the dispersion around the Brillouin zone edge K-M.

\begin{table}[!t]
\caption{The hoppings for the single orbital model of 
CuAlO$_2$ and Na$_x$CoO$_2$}
\label{table1}
\begin{tabular}{c| c c c c c c}
\hline
&\hspace{0.6cm}$t_1 [eV]$&\hspace{0.6cm}$t_2/t_1$&\hspace{0.6cm}$t_3/t_1$&\hspace{0.6cm}$t_4/t_1$&\hspace{0.6cm}$t_5/t_1$\\  
\hline 
CuAlO$_2$	&-0.45 &   0.003  & -0.069 & -0.030 & 0.034  \\
Na$_x$CoO$_2$	&  0.18 & -0.24  &   -0.15 &  -0.001 & 0.008  \\
\hline
\end{tabular}
\end{table}

\subsection{The origin of the difference between CuAlO$_2$ and Na$_x$CoO$_2$}
In this section, we discuss the origin of the difference in the 
second neighbor hopping between CuAlO$_2$ and Na$_x$CoO$_2$. To see this, 
we now construct models of these materials where all of the five 
$3d$ orbitals are considered explicitly. 
Here, the difference between the $d_{3z^2-r^2}$ orbital in the five orbital 
model and the $d_{3z^2-r^2}$ orbital mentioned 
in the previous sections should be noted. 
The $d_{3z^2-r^2}$ Wannier orbital in the previous sections consists  not 
only of the $d_{3z^2-r^2}$ in the five orbital sense, but also of the 
other hybridized orbitals as well. 
In other words, the $d_{3z^2-r^2}$ Wannier orbital of the 
single orbital model considered in the previous sections takes into account 
the effect of other orbitals implicitly. Since the band structure in the 
vicinity of the Fermi level is the same between the 
single and the five orbital models, the two models give the same 
transport properties.

\begin{figure*}[htbp]
\includegraphics[width=0.8\hsize]{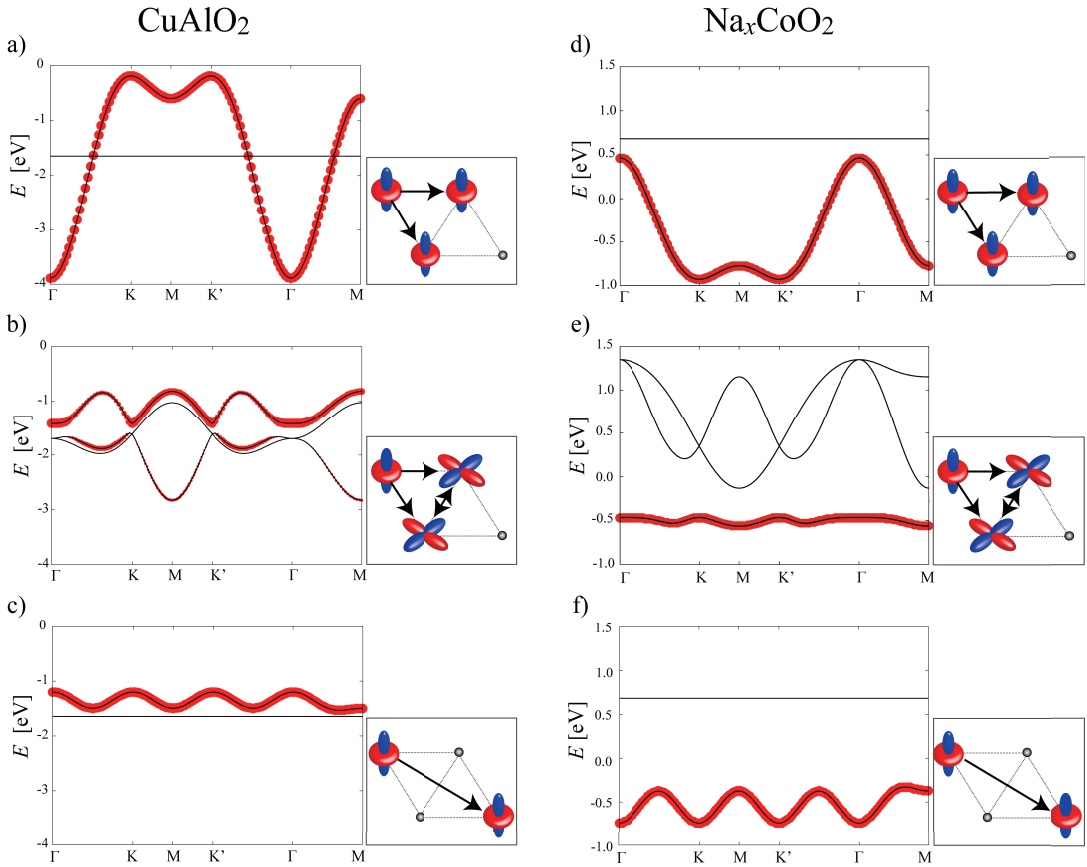}
\caption{
Band dispersion of models constructed by extracting some of the orbitals
and hoppings from the $3d$ five orbital models of CuAlO$_2$((a)-(c)) and 
Na$_x$CoO$_2$((d)-(e)) The thickness of the lines represent the 
strength of the $d_{3z^2-r^2}$ orbital character.
(a)(d) Only the $d_{3z^2-r^2}$ orbital extracted, only the nearest neighbor 
hopping considered. 
(b)(e) $d_{3z^2-r^2}$ and $d_{xy/x^2-y^2}$ orbitals extracted, only the 
nearest neighbor interorbital hopping considered.
(c)(f) Only the $d_{3z^2-r^2}$ orbital extracted, 
only the second nearest neighbor hopping considered. 
}
\label{fig6}
\end{figure*}  

In the five orbital model, the most relevant band originates from the 
$d_{3z^2-r^2}$ orbital\cite{comment} in both CuAlO$_2$ and Na$_x$CoO$_2$,  
but other orbitals can have contribution to the 
band shape due to the hybridization. To see this effect,
we first hypothetically vary the 
on-site energy of the orbitals other than $d_{3z^2-r^2}$.
We find that the $d_{xz/yz}$ orbitals have small effect on the band shape 
in both of the materials. On the other hand, we find that varying the 
on-site energy of the $d_{xy/x^2-y^2}$ orbitals affects the flatness of the 
band top only in CuAlO$_2$. This means that there is a hybridization 
between $d_{3z^2-r^2}$ and $d_{xy/x^2-y^2}$ orbitals which plays an 
important role in producing the pudding mold type band in CuAlO$_2$.

\begin{figure}[htbp]
\includegraphics[width=8cm]{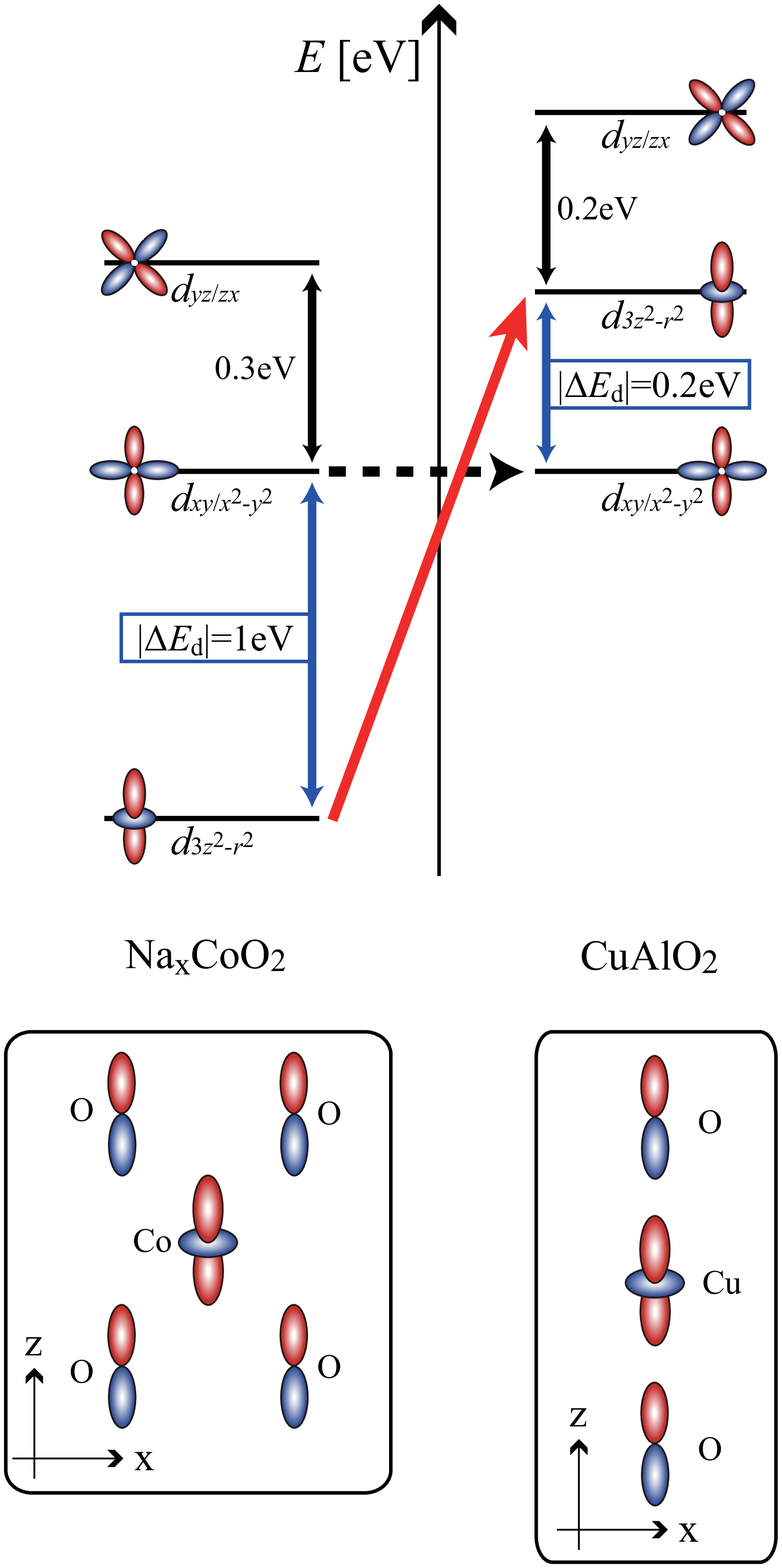}
\caption{The energy levels of the five orbital model of 
Na$_x$CoO$_2$ (left) and  CuAlO$_2$ (right). The bottom figure shows the 
difference in the ligand position between the two materials.}
\label{fig7}
\end{figure}

To further investigate this point, we consider the effects of the 
intra and interorbital hoppings step by step. In Fig.\ref{fig6}(a) and (c), 
we show the band dispersion in which only  the $d_{3z^2-r^2}$ orbital is 
extracted from the five $3d$ orbitals, and only the nearest neighbor 
(a) or the second nearest neighbor hopping (c) is considered. 
In the band dispersion shown in Fig.\ref{fig6}(b), 
we extract three orbitals out of five, 
namely, $d_{3z^2-r^2}$ and $d_{xy/x^2-y^2}$, and consider only the 
nearest neighbor interorbital hoppings between $d_{3z^2-r^2}$ and $d_{xy/x^2-y^2}$
orbitals. Here, the thickness of the lines represents the 
magnitude of the $d_{3z^2-r^2}$ orbital weight. This result shows 
that the effect of the interorbital $d_{3z^2-r^2}$-$d_{xy/x^2-y^2}$ 
orbital hopping pushes up the $d_{3z^2-r^2}$ band top around the M point. 
This effect is taken into account as a positive contribution to $t_2/t_1$ 
in the single orbital model which considers the effect of 
$d_{xy/x^2-y^2}$ implicitly. Namely, the hopping path 
$d_{3z^2-r^2}\rightarrow d_{xy/x^2-y^2}\rightarrow d_{3z^2-r^2}$ 
gives a positive contribution to the effective $t_2/t_1$ 
between second neighbor 
$d_{3z^2-r^2}$ orbitals. On the other hand, there is a negative contribution 
to $t_2/t_1$ of the single orbital model  coming from the direct 
second nearest neighbor $d_{3z^2-r^2}$-$d_{3z^2-r^2}$ 
hopping in the five orbital model, 
which pushes down the band around the M point, 
as shown in Fig.\ref{fig6}(c). In CuAlO$_2$, the positive and negative 
contributions almost cancel with each other, resulting in a 
very small $t_2/t_1$ in the single orbital model.

A similar five orbital analysis for Na$_x$CoO$_2$, shown in 
Figs.\ref{fig6}(d)-(f), reveals that the situation is different.
From Fig.\ref{fig6}(e), it can be seen that the dispersion 
originating from the $d_{3z^2-r^2}$-$d_{xy/x^2-y^2}$ hybridization 
is very small, namely, the contribution of 
the $d_{3z^2-r^2}\rightarrow d_{xy/x^2-y^2}\rightarrow d_{3z^2-r^2}$ path 
to $t_2$ in the single orbital model is negligible. On the 
other hand, there is a direct $d_{3z^2-r^2}$-$d_{3z^2-r^2}$ 
second neighbor hopping, which gives a positive contribution to 
$t_2/t_1$ in the single orbital model, pushing up the band 
around the M point. Hence, in Na$_x$CoO$_2$, $t_2/t_1$ is 
large and the flatness of the band does not appear around 
K-M-K.

The origin of the 
difference between the two materials in the 
contribution of $d_{3z^2-r^2}\rightarrow d_{xy/x^2-y^2}\rightarrow d_{3z^2-r^2}$
path comes from the difference in the energy level offset between 
$d_{3z^2-r^2}$ and $d_{xy/x^2-y^2}$ in the five orbital model 
${\Delta}E_d=E(d_{xy/x^2-y^2})-E(d_{3z^2-r^2})$. As shown in 
Fig.\ref{fig7}, this offset is large in Na$_x$CoO$_2$, while 
relatively small in CuAlO$_2$. A smaller energy difference 
gives larger contribution of the 
$d_{3z^2-r^2}\rightarrow d_{xy/x^2-y^2}\rightarrow d_{3z^2-r^2}$ path.
The difference in ${\Delta}E_d$ can be understood from 
the lattice structure. As shown in the bottom of Fig.\ref{fig7}, 
the oxygen atoms in CuAlO$_2$ are located at positions toward which 
the $d_{3z^2-r^2}$ orbitals are elongated, while that is not the 
case for Na$_x$CoO$_2$. Therefore, the crystal field of the ligand atoms 
pushes up the $d_{3z^2-r^2}$ level, locating it just above the 
$d_{xy/x^2-y^2}$ level.

\section{Conclusion}

To conclude, CuAlO$_2$ is a very good candidate for 
thermoelectric material with large Seebeck coefficient 
coexisting with large conductivity. The origin of this is the 
pudding mold band similar to, but different from, the one in 
Na$_x$CoO$_2$. In CuAlO$_2$, the negligibly small 
second nearest neighbor hopping does not affect the 
flat portion of the band around the Brillouin zone edge 
already present in the nearest 
neighbor hopping model on the triangular lattice. 
Moreover, the additional presence 
of the third (and the fifth) hopping integrals makes 
the band even more flat around the Brillouin zone edge.
This is in sharp contrast with the case of Na$_x$CoO$_2$, 
where the large second nearest neighbor hopping moves the 
flat portion of the band to the $\Gamma$ point area.
The difference of the band shape between the two materials 
comes from the difference in the 
$d_{3z^2-r^2}\rightarrow d_{xy/x^2-y^2}\rightarrow d_{3z^2-r^2}$ path 
in the five orbital sense ; a fairly close energy levels between 
$d_{3z^2-r^2}$ and  $d_{xy/x^2-y^2}$ gives rise to a large contribution 
from this path, which cancels out the direct $d_{3z^2-r^2}$-$d_{3z^2-r^2}$ 
contribution to the second neighbor hopping.
The difference in the energy difference between 
$d_{3z^2-r^2}$ and  $d_{xy/x^2-y^2}$ levels comes from the position 
of the oxygen atoms. 
A large Seebeck coefficient comparable to the model of Na$_x$CoO$_2$, 
despite a much larger band width, implies that the band shape is 
extremely ideal, and a very good thermoelectric properties,
especially a large power factor,  are 
expected once large amount of hole doping is accomplished.

\section{Acknowledgement}
We would like to thank M. Nohara for fruitful discussions.  
Part of the calculation has been done on the computer facilities, 
ISSP, University of Tokyo. H.S. acknowledges support from JSPS. (Grants No. 23009446)


\begin{thebibliography}{99}
\bibitem{Mahan} For a general review on the theoretical 
aspects as well as experimental observations 
of thermopower, see,  G.D. Mahan {\it Good Thermoelectrics, 
Solid State Physics {\bf 51}, 81 (1997).}
\bibitem{Terasaki} I. Terasaki, Y. Sasago and K. Uchinokura, Phys. Rev. B {\bf 56} R12685 (1997).

\bibitem{Koshibae}
W. Koshibae, K. Tsutsui and S. Maekawa, Phys. Rev. B {\bf 62} 6869 (2000).
\bibitem{KoshibaePRL} W. Koshibae and S. Maekawa, Phys. Rev. Lett. {\bf 87}, 236603 (2001).
\bibitem{Singh}
D.J. Singh, Phys. Rev. B {\bf 61}, 13397 (2000).

\bibitem{Kuroki} K.Kuroki and R.Arita, J.Phys. Soc.Jpn. {\bf 76} 083707 (2007).
\bibitem{Kuriyama} H. Kuriyama, M. Nohara, T. Sasagawa, K. Takubo, 
T. Mizokawa, K. Kimura, and H. Takagi,   in Proc. 25th International Conference on Thermoelectrics  (IEEE, Piscataway, 2006).
\bibitem{Shibasaki2} S.Shibasaki, W.Kobayashi, and I.Terasaki: Phys. Rev. B {\bf 74}, 235110 (2006).
\bibitem{Usui} H. Usui, R. Arita, and K. Kuroki, Journal of Physics. Condensed Matter : an Institute of Physics Journal {\bf 21}, 064223 (2009).
\bibitem{Okamoto} Y. Okamoto, S. Niitaka, M. Uchida, T. Waki, M. Takigawa, Y.
Nakatsu, A. Sekiyama, S. Suga, R. Arita, and H. Takagi, Phys.
Rev. Lett. {\bf 101}, 086404 (2008).
\bibitem{Arita} R. Arita, K. Kuroki, K. Held, A. V. Lukoyanov, S. Skornyakov, and V. I. Anisimov, Phys. Rev. B {\bf 78}, 115121 (2008).
\bibitem{PSun} P. Sun, N. Oeschler, S. Johnsen, B.B. Iversen, and F. Steglich, Appl. Phys. Express {\bf 2} 091102 (2009).
\bibitem{UsuiFeAs} H. Usui, K. Kuroki, S. Nakano, K. Kudo, and M. Nohara, 
arXiv:1211.7176.
\bibitem{Wissgott} P. Wissgott, A. Toschi, H. Usui, K. Kuroki, and K. Held, 
Phys. Rev. B {\bf 82}, 201106 (2010).
\bibitem{Wissgott2} P. Wissgott, A. Toschi, G. Sangiovanni, and K. Held
Phys. Rev. B {\bf 84}, 085129 (2011).
\bibitem{UsuiNaxCoO2} H. Usui, PhD Thesis, 2012, Univ. of 
Electro-Communications, unpublished.
\bibitem{Yoshida} H. Katayama-Yoshida, T. Koyanagi, H. Funashima, H. Harima, A. Yanase, Solid State Comm.{\bf 126}, 135 (2003).
\bibitem{Poopanya} P. Poopanya, A. Yangthaisong, C. Rattanapun, A. Wichainchai, J. of Electronic Mat., {\bf 40} 987 (2011).
\bibitem{Nakanishi} A. Nakanishi, H. Katayama-Yoshida, Solid State Comm.{\bf 152}, 24 (2012).
\bibitem{wien2k}
P. Blaha, K. Schwarz, G.K.H. Madsen, D. Kvasnicka, and J. Luitz, 
{\it Wien2k: An Augmented Plane Wave} + {\it Local Orbitals Program for Calculating Crystal Properties} (Vienna University of Technology, Wien, 2001).
\bibitem{PBE} J. P. Perdew , K. Burke, and M. Ernzerhof, 
Phys. Rev. Lett. {\bf 77}, 3865 (1996).
\bibitem{wannier} N. Marzari and D. Vanderbilt, Phys. Rev. B 
{\bf 56}, 12847 (1997);  
I. Souza, N. Marzari, and D. Vanderbilt, Phys. Rev. B {\bf 65}, 035109 (2001).
The Wannier functions are generated by the code developed by
A. A. Mostofi, J. R. Yates, N. Marzari, I. Souza, and D. Vanderbilt,
(http://www.wannier.org/).  
\bibitem{w2w} J. Kunes, R. Arita, P. Wissgott, A. Toschi, H. Ikeda, and K. Held, Comp. Phys. Commun. {\bf 181} 1888 (2010).
\bibitem{comment} For Na$_x$CoO$_2$, the following notation is 
more commonly used\cite{Singh}. The $3d$ orbitals consist of 
the low-lying $t_{2g}$  and the high-lying $e_g$ orbitals, 
and the $t_{2g}$ orbitals are split into $a_{1g}$ and 
$e_{g}'$ orbitals. In the present notation, 
the $d_{3z^2-r^2}$ orbital corresponds to the 
commonly used $a_{1g}$ orbital. We will call this the 
``$d_{3z^2-r^2}$ orbital'' in the sense that
it is the maximally localized Wannier orbital obtained 
by projecting onto the $d_{3z^2-r^2}$ orbital.
The bonding and antibonding states between 
the present $d_{xy/{x^2-y^2}}$ and $d_{xz/yz}$ orbitals 
correspond to the $e_g'$ and $e_g$ orbitals, respectively.
\end{thebibliography}
\end{document}